\documentclass[12pt,twoside]{article}
\usepackage{fleqn,espcrc1}

\usepackage{graphicx}

\newcommand{\beq}{\begin{equation}} \newcommand{\eeq}{\end{equation}}
\newcommand{\vect}[1]{\mbox{\boldmath${#1}$}} 
\newcommand{\vev}{{\vect v}}

\newcommand{\vebt}{{\vect \beta}}
\newcommand{\vep}{{\vect p}}
\newcommand{\beqa}{\begin{eqnarray}} \newcommand{\eeqa}{\end{eqnarray}}

\newcommand{\AmS}{{\protect\the\textfont2
  A\kern-.1667em\lower.5ex\hbox{M}\kern-.125emS}}

\title{A calculation of the transport coefficients of \\
 hot and dense hadronic matter \\
based on the event generator URASiMA}

\author{
N. Sasaki\address{Department of Physics, Hiroshima University,\\ 
Higashi-Hiroshima 739-8526, Japan}, 
O. Miyamura\address{Department of Physics, Hiroshima University,\\ 
Higashi-Hiroshima 739-8526, Japan}, 
S. Muroya\address{Tokuyama Women's College, Tokuyama, 745-8511, Japan}\thanks{the presenter at this conference}
and 
C. Nonaka\address{Department of Physics, Hiroshima University,\\ 
Higashi-Hiroshima 739-8526, Japan}
}
       
\begin{document}

\maketitle

\begin{abstract}
We evaluate thermodynamical quantities and the transport coefficients of a dense
and hot hadronic matter based on the event generator URASiMA
(Ultra-Relativistic AA collision Simulator based on Multiple
Scattering Algorithm) with periodic boundary conditions.
As the simplest example of the transport coefficients we investigate 
 the temperature dependence 
and the chemical potential dependence of the baryon diffusion constant of a
dense and hot hadronic matter.
\end{abstract}

\section{ Introduction} \label{intr}

Physics of a high density and high temperature hadronic matter has been
highly studied in the context of both high energy nuclear collisions and
cosmology\cite{QM99}. In recent ultra-relativistic
nuclear collisions, though the main purpose should be confirmation of
Quark-Gluon Plasma(QGP) state, physics of hot and/or dense hadronic state
dominates the system. Hence, thermodynamical properties and transport
coefficients of a hadronic matter are essentially important understand 
the space-time evolution of the exited region. 
In the cosmology, physics of hot hadrons is important not only for the 
global evolution of the early universe, but also in the nucleosynthesis problem 
baryon diffusion would play an important roll.

Because of the highly non-perturbative property of a hot and dense hadronic
state, the thermodynamical properties and transport
coefficients have not been thoroughly investigated.  
In this paper, we evaluate the transport coefficients by using statistical
ensembles generated by Ultra-Relativistic A-A collision simulator based on
Multiple Scattering Algorithm (URASiMA). Originally, the URASiMA is an event
generator for the nuclear collision experiments based on the Multi-Chain
Model(MCM) of the hadrons\cite{Kumagai}.
We improved the URASiMA to recover {\it detailed balance} at 
temperature below two hundred MeV, and we can obtain natural thermodynamical
behavior without
 Hagedorn-type strange temperature saturation. This is the first calculation of
the transport coefficient of a hot and dense hadronic matter based on
an event generator\cite{Sasaki1}.

\section{URASiMA for Statistical Ensembles} \label{URA}

The URASiMA is a relativistic event generator based on hadronic multi-chain
model, which aims at describing nuclear-nuclear collision by the superposition of
hadronic
collisions. Hadronic 2-body interactions are fundamental building blocks of
interactions in the model, and all parameters are so designed to reproduce
experimental data of hadron-hadron collisions.
Originally, the URASiMA contains
2-body process (2 incident particle and 2 out-going particles), decay
process
(1 incident particle and 2 out-going particles), resonance (2 incident
particles and 1 out-going particle) and production process (2 incident
particles and n ($\ge$ 3) out going particles).

The production process is very important for the description of the multiple
production at high energies. On the other hand, in the generation of 
statistical ensembles in equilibrium,
detailed balance between processes is essentially important. If 
re-absorption process does not exist, one-way conversion of energy into particle
production  occurs rather than heating up and results in artificial saturation of the 
temperature\cite{RQMD}.
Therefore, role of re-absorption processes is very important
and we should take it into account.
However exact inclusion of multi-particle re-absorption processes is very
difficult.
In order to treat them effectively,
multi-particle productions and absorptions are treated as 2-body
processes including resonances with succeeding decays and/or preceding
formations of the resonances. Here two body decay and formation of
resonances are assumed. For example, $NN \to NN\pi$ is described as
$NN \to NR$ followed by decay of $R \to
N \pi$, where $R$ denotes resonance.
The reverse process of it is easily taken into account. In this approach,
all the known inelastic cross-sections for baryon-baryon interactions up to
$\sqrt{s} < 3$GeV, are reproduced.

For a higher energy, $\sqrt{s} > 3$GeV, in order to give appropriate
total cross section, we need to take direct production process into account.
Only at this point, detailed balance is broken in our simulation; nevertheless,
if temperature is much smaller than 3 GeV, the influence is negligibly
small.
\begin{center}
\begin{table}[htb]
  \caption{
    Baryons, mesons and their resonances included in the URASiMA.
  }
  \label{tab:plst}
  \begin{tabular}{ccccccccc}
\hline 
    nucleon  & $N_{938} $ & $N_{1440}$ & $N_{1520}$ & $N_{1535}$ &
               $N_{1650}$ & $N_{1675}$ & $N_{1680}$ & $N_{1720}$ \\
    $\Delta$ & $\Delta_{1232}$ & $\Delta_{1600}$ &
               $\Delta_{1620}$ & $\Delta_{1700}$ &
               $\Delta_{1905}$ & $\Delta_{1910}$ &
               $\Delta_{1950}$ & \\ 
    meson & $\pi$ & $\eta$ & $\sigma_{800}$ & $\rho_{770}$ &&&& \\
\hline
  \end{tabular}
\end{table}
\end{center}
In order to obtain an equilibrium state, we put the system in a box and imposed
a periodic condition to the URASiMA as the space-like boundary condition. Initial
distributions of particles were given by uniform random distribution of
baryons in a phase space. Total energy and
baryon number in the box were fixed at initial time and conserved through-out
the simulation.  Though initial particles were only baryons, many mesons were
produced through interactions. After a thermalization time-period of about
100 fm/c, the system seemed to be stationary.
In order to confirm the achievement of equilibrium, we calculated energy
distributions and particle numbers. Slope parameters of energy distribution
of all particles became the same value in the accuracy of statistics.
Thus, we may call this value as the temperature of the system. The fact that
numbers of species saturate indicates the achievement of chemical
equilibrium\cite{Sasaki1}. 

Running the URASiMA many times with the same total energy
and total baryons in the box and taking the stationary configuration later
than $t=150$ fm/c, we obtained statistical ensemble with fixed
temperature and fixed baryon number(chemical potential).
By using the ensembles obtained through above mentioned manner, we can
evaluate thermodynamical quantities and equation of states\cite{Sasaki2}.

\section{Diffusion Constant} \label{Trans}

According to the Kubo's Linear Response Theory, the correlation of the
currents stands for admittance of the system(first fluctuation dissipation
theorem) and equivalently, random-force correlation gives impedance(Second
fluctuation dissipation theorem) \cite{Kubo}. As the simplest example, we
here focus our discussion on the diffusion constant. First
fluctuation dissipation theorem tells us that diffusion constant $D$ is
given by current(velocity) correlation,
\beq
D =\frac{1}{3} \int_{0}^{\infty}<\vev(t)\cdot \vev(t+t')> dt'. \label{eqn;fdt}
\eeq
If the correlation decreases exponentially, i.e., $
<\vev(t)\cdot \vev(t+t')> \propto \exp{(- \frac{t'}{\tau})}, \label{eqn;rlx}
$
with $\tau$ being relaxation time,
diffusion constant can be rewritten in a simple form, \beq
D = \frac{1}{3}<\vev(t)\cdot \vev(t)> \tau .\label{eqn;difcon} \eeq
Because of the
relativistic nature of our system, we should use $\vebt = \frac{\vev}{c} =
\frac{\vep}{E}$ instead of $\vev$ in eq.(\ref{eqn;fdt}) and $D$ is obtained
by,
\beqa
D &=& \frac{1}{3}\int_{0}^{\infty}<\vebt(t)\cdot \vebt(t+t')> dt' {c^2},
\label{eqn;fdt2}
\\
&=&\frac{1}{3}<\vebt(t)\cdot \vebt(t)> c^2 \tau ,\label{eqn;difcon2} \\
&=&\frac{1}{3}<\left(\frac{\vep(t)}{E(t)}\right)\cdot
\left(\frac{\vep(t)}{E(t)}\right)> c^2 \tau ,\eeqa
with $c$ being the velocity of light.
Figure 1 shows correlation function of the velocity of baryons. The figure
indicates that exponential damping is very good approximation.
Figure 2 displays the our results of baryon diffusion constant in a hot and
dense hadronic matter.
\begin{figure}[htbp]
\begin{center}
  \includegraphics[scale=0.60 
]{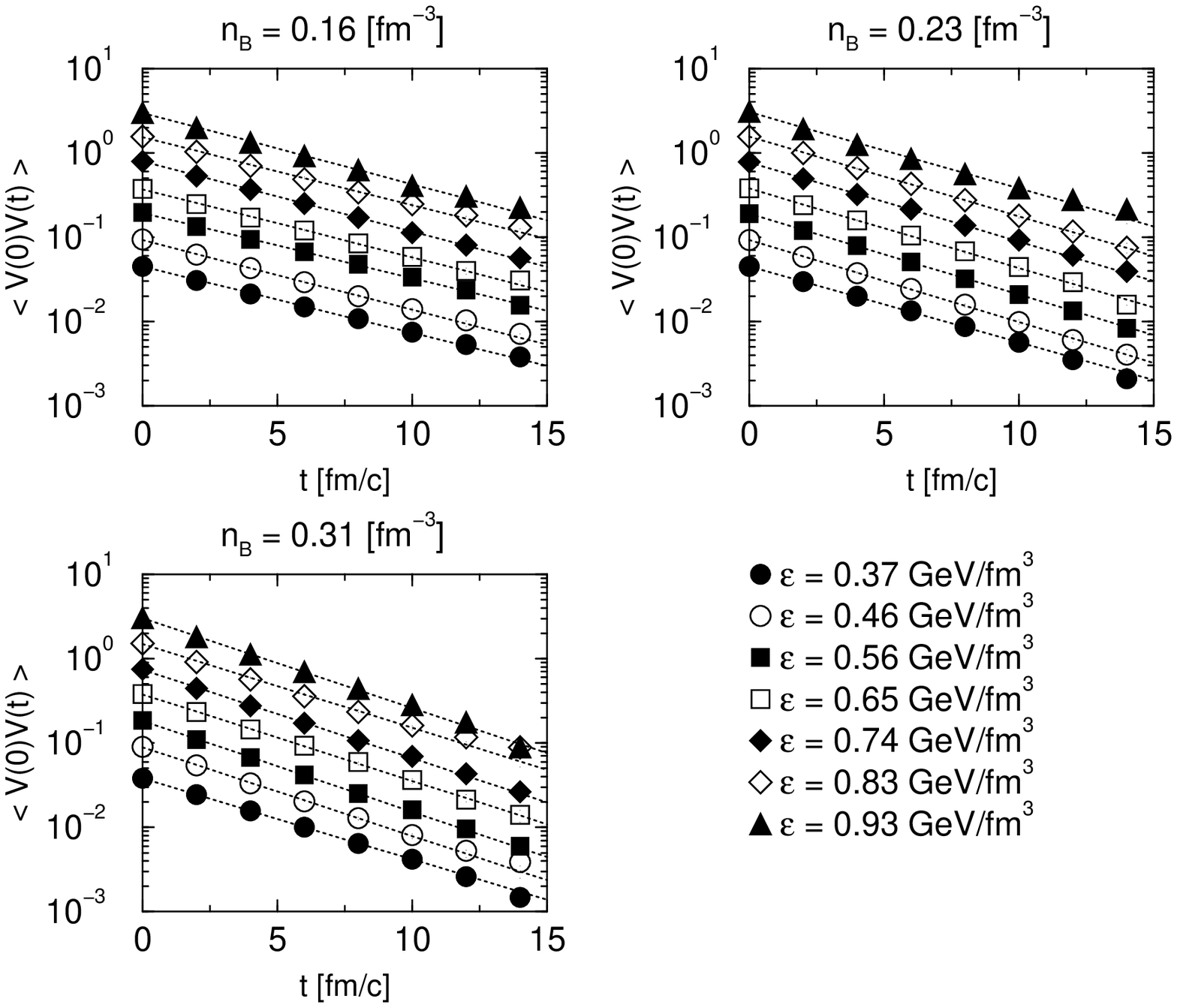} 
  \caption{
    Velocity correlation of the baryons as a function of time.
    Lines correspond to the fitted results by exponential function.
    Normalizations of the data are arbitrary.
}
\end{center}

\begin{center}
  \includegraphics[scale=0.50]{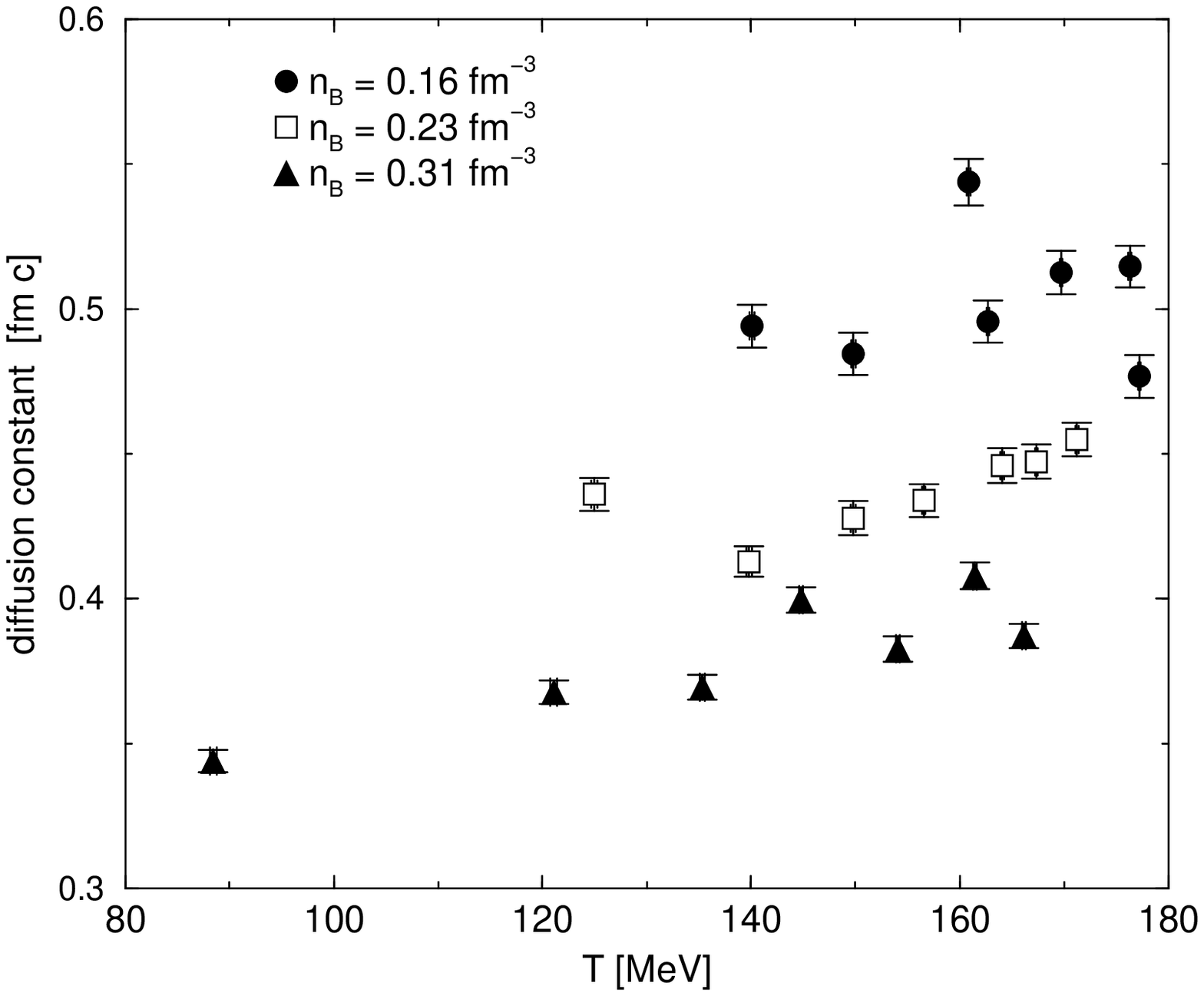}
  \caption{ Diffusion constant of baryons.}
\end{center}
\end{figure}

Our results show clearer dependence on the baryon number density while
dependence on temperature is mild. This result means that, for the 
random walk of the baryons in our system,
baryon-baryon collision process is more important than baryon-meson collisions.
In the inhomogeneous big-bang nucleosynthesis scenario, 
baryon-diffusion plays an important roll.
The leading part of the scenario is played by the difference
between proton diffusion and neutron diffusion\cite{inho}. In our
simulation,
strong interaction
dominates the system and we assume charge independence in the strong
interaction, hence, we can not discuss difference between proton and
neutron.
However obtained diffusion constant of baryon in our simulation can give
some
kind of restriction to the diffusion constants of both proton and neutron.

\section{Concluding Remarks} \label{conc}

We evaluate diffusion constants of baryons in the hot and dense hadronic
matter with use of statistical ensembles obtained by an event generator URASiMA.
Our results show clear dependence on baryon number density and
weak dependence on temperature.
The temperature in our simulation is limited to a small range, i.e., from
100 MeV to 200 MeV, and this fact can be one of the reasons  for unclear 
dependence on temperature.
Strong baryon number density dependence indicates that, for the baryon
diffusion process, baryon plays more important roll than light mesons. In this
sense our simulation corresponds to high density region and a non-linear
diffusion process occurs. 

Calculation of the diffusion constants is the
simplest example of first fluctuation dissipation theorem and the entrance 
of the non-equilibrium physics.
\begin{figure}[tbh]
\begin{center}
  \includegraphics[scale=0.5]{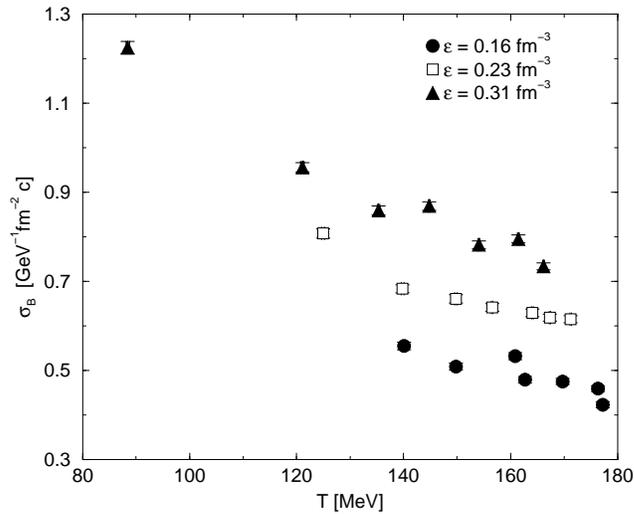}\\
\caption{Baryon number charge conductivity as a function of temperature}
\end{center}
\end{figure}
From diffusion constant, we can calculate charge
conductivity\cite{textbook}.
Figure 3 shows baryon number conductivity $\sigma_{\rm B}$, \beq
\sigma_{\rm B} = \displaystyle{\frac{n_{\rm B}}{k_{\rm B}T}}D, \eeq
where $n_{\rm B}$ is baryon number density, $T$ is temperature and $k_{\rm
B}$ is Boltzmann constant(put as unity through out this paper),
respectively.
Therefore, if we want, we can discuss Joule heat and entropy production in
the
{\it Baryonic circuit} with use of the above baryon number conductivity.

In principle, taking correlation of appropriate currents, i.e. energy
flow, baryon number current, stress-tensor, etc.,
we can evaluate any kinds of transport
coefficients. Though, in relativistic transport theory, there exist several delicate
points\cite{Namiki},  once we establish
phenomenological equations for the high temperature and high density
hadronic matter, we can evaluate the appropriate transport coefficients in the same
manner. Detailed discussion will be reported in our forthcoming
paper.

\noindent
{\bf Acknowledgment}

The authors would like to thank prof. M.\ Namiki and prof. T.\ Kunihiro for their
 fruitful comments. 
This work is supported by Grant-in-Aid for scientific research number
11440080 of Ministry of
Education, Science, Sports and Culture,
Government of Japan (Monbusho).  Calculation has been done at Institute for Nonlinear
Sciences and Applied Mathematics, Hiroshima University.


\end{document}